\documentclass[]{interact}
\usepackage{xcolor}
\usepackage[normalem]{ulem}
\usepackage{epstopdf}
\usepackage[caption=false]{subfig}

\usepackage[numbers,sort&compress]{natbib}
\bibpunct[, ]{[}{]}{,}{n}{,}{,}
\renewcommand\bibfont{\fontsize{10}{12}\selectfont}
\makeatletter
\def\NAT@def@citea{\def\@citea{\NAT@separator}}
\makeatother

\theoremstyle{plain}

\theoremstyle{definition}

\theoremstyle{remark}

\begin{document}


\title{
Calculation 
of the
local environment of 
a 
barium monofluoride 
molecule in an
argon matrix:
A step towards
using 
matrix-isolated
BaF
for determining 
the electron
electric dipole
moment
}

\author{
\name{
Ricardo L. Lambo\textsuperscript{a}
and
Gregory K. Koyanagi\textsuperscript{a}
and
Anita Ragyanszki\textsuperscript{a}
and
Marko Horbatsch\textsuperscript{a}
and
Rene Fournier\textsuperscript{a}
and
Eric A. Hessels\textsuperscript{a}\thanks{CONTACT E.~A. Hessels Email: hessels@yorku.ca}}
\affil{\textsuperscript{a}York University, 
Toronto, Ontario, Canada (EDM$^3$ Collaboration)}
}

\maketitle

\begin{abstract}
The local environment 
of a 
barium monofluoride
(BaF)
molecule
embedded in an argon
matrix 
is calculated.
A substitution of 
a 
BaF
molecule for 
four 
Ar 
atoms is found to 
be 
strongly favoured
compared to substitutions
for other numbers of 
Ar
atoms.
The 
equilibrium positions of the 
BaF 
molecule
and
its nearby 
Ar
neighbours
are 
found
by minimizing the total energy.
The potential barrier 
that prevents the migration
of the 
BaF 
molecule within the 
solid 
and
the barrier
that prevents its rotation
are calculated.
At the cryogenic temperatures
used by the 
EDM$^3$
collaboration,
these barriers
are sufficiently large 
to fix the position
and orientation of the molecule.
Knowledge of the local environment
of matrix-isolated 
BaF
molecules
is 
essential for 
the 
EDM$^3$ 
collaboration,
which is using 
them in
a precision 
measurement of
the electron
electric dipole
moment.
\end{abstract}
\resizebox{25pc}{!}{\includegraphics{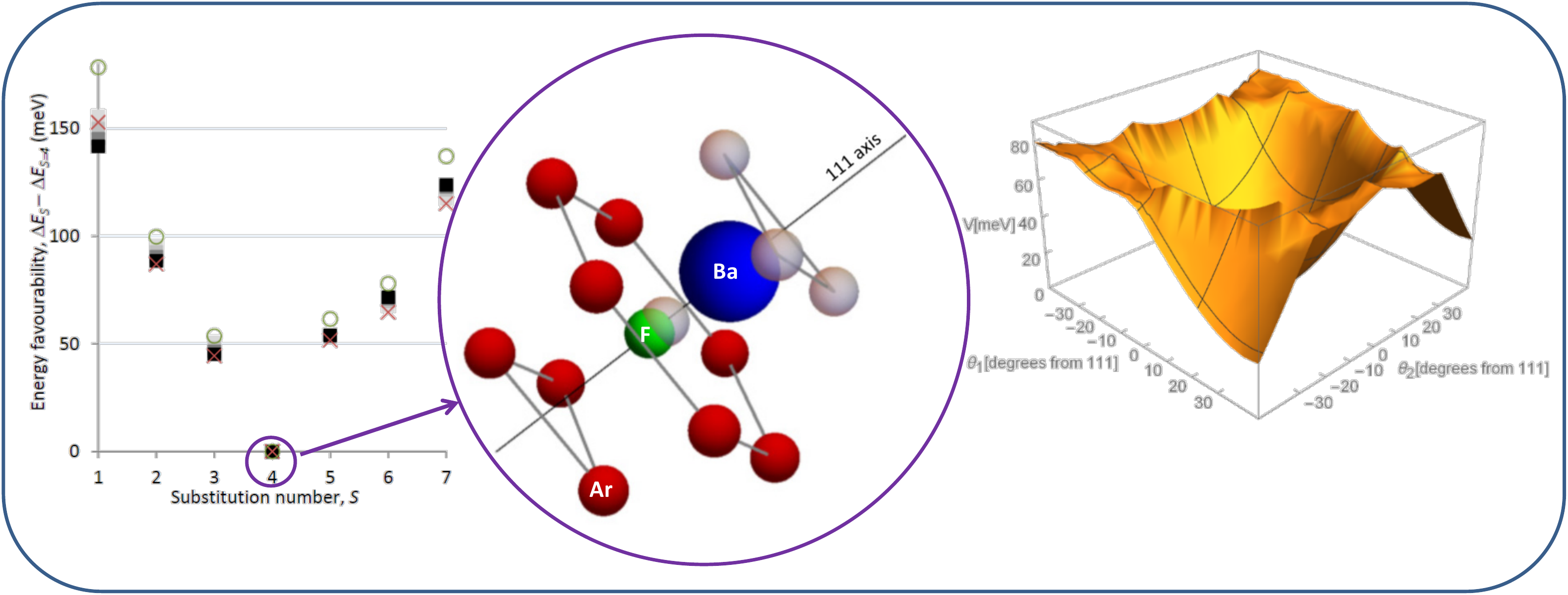}}

\begin{keywords}
Matrix isolation; 
barium monofluoride; 
argon
\end{keywords}

\section{Introduction}

Barium 
monofluoride
embedded in a 
solid 
argon matrix
is being used
\cite{vutha2018orientation}
by the 
EDM$^3$
collaboration for 
pursuing a measurement
of the 
electron electric 
dipole moment
(eEDM).
Current measurements
\cite{acme2018improved}
of the 
eEDM
already 
put strong limits on possible
beyond-the-standard-model
physics
(for energy scales
of up to 
100~TeV)
that would lead to 
the level of
time-reversal (T) violation 
required to understand the 
asymmetry between matter and
antimatter in the universe.
Measurements of the 
eEDM 
at higher levels of accuracy
will test 
T-violating
physics at even
higher energy scales.

In this work,
we calculate 
the local environment 
of a 
BaF
molecule 
embedded in 
an 
Ar
matrix.
A substitution of 
a 
BaF
molecule for 
four
Ar
atoms 
is found to be 
strongly favoured
energetically 
compared to substitutions
for other numbers of 
Ar
atoms.
The equilibrium 
positions of the
BaF
molecule 
and of the neighbouring
Ar
atoms 
are calculated.
Most importantly
for the 
work of the 
EDM$^3$
collaboration,
the potential barriers
that prevent the 
BaF 
molecule
from migrating through
the solid 
and from 
rotating within
the solid are determined.
These barriers will
allow for a large 
sample of stationary,
non-rotating
BaF
molecules
(without the need
for an external 
electric field
to inhibit 
their rotations
\cite{vutha2018orientation}).

In a previous work
\cite{koyanagi2022accurate},
we calculated 
the 
ground-state
energies of the 
BaF-Ar 
triatomic
system using 
high-precision
all-electron
relativistic
quantum-mechanical
calculations 
(using a 
scalar-relativistic 
approach within the framework of a 
second-order
Douglas-Kroll-Hess
approach)
that
include 
correlation
and that are 
extrapolated to 
the complete basis
set limit.
These energies were
calculated for a large
range of positions 
(angles
$\theta$
and
distances
$r$)
of the 
Ar
atom
relative to the
BaF
molecule
using the
CCSD(T)
method.
The calculations provide
a smoothly varying 
potential energy 
versus
$r$
and
$\theta$
for the 
interaction between
BaF
and 
Ar.
This potential
is used here,
along with 
the 
well-known
Ar-Ar
interatomic 
potential to
calculate the 
local environment 
near the embedded
BaF
molecule.

Recently,
BaF
molecules
have been embedded in both
Ne
\cite{li2022baf}
and
Ar
\cite{corriveau2022baf}
cryogenic solids
by the 
EDM$^3$
collaboration.
In both cases,
laser-induced
fluorescence 
is observed.
Earlier work
\cite{knight1971hyperfine}
also studied
matrix-isolated
BaF
molecules,
including 
optical absorption
and 
electron spin resonance
studies.
Knowledge of the 
position of the 
BaF 
molecule
within the 
fcc 
Ar
crystal
is needed  
to calculate
shifts 
of
the
BaF
energy levels
due to the 
Ar
matrix
and interpret
the observed
spectra.
The present work 
will also allow
for future calculations
of the 
oscillatory 
modes 
(e.g., 
librational motion
and 
centre-of-mass
oscillatory motion)
for the 
matrix-isolated 
BaF
molecule.
An understanding of the 
local environment of the 
BaF
molecule within an
Ar
solid
will help to
guide continuing
work of the 
EDM$^3$
collaboration.

\section{
Methods
\label{sec:methods}}

In this work, 
the geometry and energy of 
a 
BaF
molecule 
in an
Ar
crystal is determined.
An ideal 
Ar 
crystal
forms a
face-centred-cubic 
(fcc)
structure with a cube 
of size 
$a=5.3118$~\AA~\cite{barrett1964x}.
Relative to one
Ar
atom, 
there are 
$n_1=12$ 
nearest neighbours
at
$b_1=a/\sqrt{2}$,
with subsequent 
sets 
(of size
$n_k$)
of nearest 
neighbours
at 
$b_k=\sqrt{k} b_1$
(for 
subsequent
integer values 
$k$).
Our simulations start
with a 
cluster of 
Ar
atoms
within a sphere of
radius 
$b_n$.
In the same manner as
was done in 
Ref.~\cite{tao2015heat},
the
outer part of this sphere 
(those farther than  
$b_m$
from the centre)
have their positions 
fixed at the ideal
Ar
fcc 
crystal positions.
Inside this shell,
a single 
BaF
molecule 
is situated near the 
centre,
surrounded by 
$M$ 
Ar
atoms.
These 
$M$
Ar
atoms
and 
one
BaF
molecule 
are allowed to move
to minimize
the overall 
interaction energy of 
the system.

In particular,
the interaction energy is 
minimized while 
varying 
$3M+5$ 
parameters:
the positions of the 
$M$
Ar
atoms
and of 
the centre of mass
of the 
BaF molecule,
as well as
the
two angles that 
define the 
orientation of the 
BaF 
molecule.
As in 
Ref.~\cite{koyanagi2022accurate},
the 
separation between the 
Ba
and
F
nuclei 
is fixed
at 
2.16~\AA,
the separation determined 
from rotational spectroscopy
\cite{bernard1992laser}.
The
large
BaF 
binding energy 
(6 eV 
\cite{ehlert1964mass,hildenbrand1968mass}) 
compared to
the 
BaF-Ar 
binding energy 
(23 meV 
\cite{koyanagi2022accurate}),
along with the larger
equilibrium separations 
for the 
BaF-Ar
system,
leads to a much stronger 
restoring force for 
stretching this 
2.16~\AA~separation
compared to 
typical 
BaF-Ar forces,
justifying 
a 
fixed
BaF 
internuclear separation.

Two independent 
calculations 
using different methods
are employed for this 
minimization to 
verify that the 
global minimum is found.
One hundred
independent simulated annealing runs
with different random 
initial configuration 
and different cooling schedules
from a temperature
$T_{\rm high}$
(of between 
50 
and 
100~K)
to 
$T_{\rm low} < 0.1$~K
are carried out 
for each energy 
minimization performed.
Averaging the 
low-$T$ 
configurations
within a simulation 
yielded the 
lowest-energy 
configuration for that run.
The five 
lowest energies found among 
the 
100 
runs are typically within a few
meV.
The lowest of these 
is further refined 
by local minimization 
to an accuracy of
better than 
0.1~meV.
An independent 
program for minimization uses 
$10^5$ 
trials 
with randomly chosen
initial positions and 
an adaptive gradient search.
The lowest twenty energies
obtained from these trials
agree to better than
0.1~meV,
and these results
(their energies and 
positions)
agree with the annealing 
results.

The number $M$
is chosen to be 
$S$
fewer than the 
number of atoms that 
would fully occupy
the sphere
($M=1+\sum_{i=1}^m n_i-S)$,
which allows the 
BaF
molecule to substitute
for 
$S$
Ar 
atoms.
The total number of 
Ar
atoms
(including the fixed outer shell)
in the simulations is
$N=1+\sum_{i=1}^n n_i-S$.

The energy being minimized 
is the sum of 
pairwise interactions:
\begin{eqnarray}
E
=&&
\sum_{i=1}^{M}\sum_{j=i+1}^{N}
V_{\rm Ar-Ar}(|\Vec{r}_i-\Vec{r}_j|)
+
\sum_{i=1}^{N}
V_{\rm BaF-Ar}(|\Vec{r}_i-\Vec{r}_0|,\theta_i).
\label{eq:totalE}
\end{eqnarray}
Here,
$\Vec{r}_0$
is the position of the 
BaF molecule,
defined as the geometric
average of  
$\Vec{r}_{\rm Ba}$
and 
$\Vec{r}_{\rm F}$
(i.e.,
the centre point between 
the 
Ba
and 
F
nuclei),
$\Vec{r}_i$
is the position of the 
$i^{\rm th}$
Ar 
nucleus,
and 
$\theta_i$
is the angle between the 
internuclear 
axis
($\Vec{r}_{\rm F}-\Vec{r}_0$)
and 
the 
$i^{\rm th}$
Ar 
atom
($\Vec{r}_i-\Vec{r}_0$).

We have recently calculated
\cite{koyanagi2022accurate}
the
BaF-Ar
interaction
energy,
$V_{\rm BaF-Ar}(r,\theta)$,
by calculating 
the 
ground-state
energies of the 
BaF-Ar 
triatomic
system 
for 
1386 
values of 
$r$
and 
$\theta$
using 
high-precision
all-electron
relativistic
quantum-mechanical
calculations 
that
include 
correlation
and that are 
extrapolated to 
the complete basis
set limit.
These calculations 
used a 
scalar-relativistic 
approach within the 
framework of a 
second-order
Douglas-Kroll-Hess
approach.
A fit provided in that work
or,
alternatively,
interpolations
and extrapolations
provide 
$V_{\rm BaF-Ar}(r,\theta)$
for intermediate values of 
$r$
and
$\theta$.
Uncertainties from 
this calculation of 
$V_{\rm BaF-Ar}(r,\theta)$
are also provided in 
Ref.~\cite{koyanagi2022accurate}.

The interaction energy between two 
Ar
atoms,
$
V_{\rm Ar-Ar}
^{\rm pairwise}(r)
$
is precisely known
\cite{jager2009ab}.
To correctly describe 
an 
Ar
crystal, 
however,
corrections
must be included
\cite{schwerdtfeger2016towards}
to these 
pairwise 
Ar-Ar
interactions.
The dominant correction 
is due to
the
zero-point
energy
of the 
Ar 
atoms,
which requires
an averaging of 
$
V_{\rm Ar-Ar}
^{\rm pairwise}(r)
$
over the 
positional 
probability 
distributions 
that results from the 
zero-point
motion
of
the two atoms.
A second,
slightly-smaller
correction
is due to 
three-body 
Ar-Ar-Ar
interactions.
Without these corrections, 
a calculation
using only the 
two-body potential
predicts
an
fcc
crystal with the incorrect 
lattice constant 
and cohesive energy.
Following the example of
Ref.~\cite{bezrukov2019empirically},
we compensate for these effects
by using a modified potential 
\begin{equation}
V_{\rm Ar-Ar}(r)
=
\alpha V
_{\rm Ar-Ar}
^{\rm pairwise}
(\beta r).
\label{eq:scaledArAr}
\end{equation}
Coefficients 
$\alpha=0.8395$
and
$\beta=0.9815$
are chosen 
to match the 
experimental 
Ar
fcc 
cube 
dimension of 
$a=5.3118$~\AA~\cite{barrett1964x}
and cohesion energy 
per atom of 
$E_{\rm coh}=80.05$~meV
\cite{beaumont1961thermodynamic}.

Four uncertainties in these
simulations are investigated.
The first is due to the 
finite size of the cluster 
used for the calculation. 
For all simulations, 
the calculations are repeated
with increasing numbers 
$N$ 
and 
$M$
of 
Ar 
atoms
and
the convergence of
our results with 
increasing cluster
size provides 
an estimate of the
resulting uncertainty.

A second comes from the 
uncertainty in our 
calculated
BaF-Ar
potentials 
in 
Ref.~\cite{koyanagi2022accurate}.
In that work, 
we repeat our calculations
with increasing basis set 
sizes: 
$n \zeta$,
with 
$n=2$
through
$5$.
We make two extrapolations
of our results,
one from 
$n=2$,
$3$
and
$4$,
and the other
(more precise one)
from 
$n=3$,
$4$
and
$5$.
Based on comparisons to 
measured quantities in 
BaF, 
Ba,
Ba$^+$
and
Ar,
we estimate the 
uncertainty in our 
345
extrapolated
results
to be
one quarter
of the difference 
between these two 
extrapolations.
To determine the effect of 
these uncertainties on our 
present cluster simulations,
we repeat the simulations
with both the
234
and 
345
results from 
Ref.~\cite{koyanagi2022accurate}.
The uncertainty in our simulations
is expected to be 
one quarter of the difference
between the two results. 

Thirdly, 
and most importantly,
we investigate the approximation
inherent in
using the scaled
Ar-Ar
potential of 
Eq.~(\ref{eq:scaledArAr}).
To do this, 
we repeated cluster calculations
using 
$
V_{\rm Ar-Ar}
^{\rm pairwise}(r)
$
in place of 
Eq.~(\ref{eq:scaledArAr}).
These calculations led
to 
a
cohesive energy
$E_{\rm coh}$
and
Ar-Ar 
separation
$b_1$
that differ from
measured values
by a factor 
of 
$\alpha$
and 
$\beta$,
respectively.
The difference 
between the calculations
with and
without
scaling
of the 
Ar-Ar
potential, 
should give the 
scale for the 
effect of the
approximation used here.

Finally,
the fourth uncertainty 
involves  
four-atom
BaF-Ar-Ar
interactions.
To estimate this
effect,
we calculate 
energies
for the
BaF-Ar-Ar
four-atom system
for twenty
geometries
with separations 
of between
3
and 
7~\AA,
which covers the  
most important range of 
separations for
our simulated solids.
We 
compare the 
234
extrapolation of this 
binding 
energy 
to the sum of the 
Ar-Ar
binding energy
plus
the two 
BaF-Ar
contributions
(also calculated using a 
234 
extrapolation).
The difference 
between the full
BaF-Ar-Ar
calculation
and the sum of the 
two-body 
contributions
is typically less than 
one percent. 
It 
is sometimes positive 
and sometimes negative,
and gets smaller quickly
with increasing 
distances.
Therefore, 
we estimate the net effect
due to 
this 
four-atom
effect
to be less than
one percent.

\section{
Results
}
\subsection{
Number of  
A\MakeLowercase{r}
atoms substituted
for 
a 
B\MakeLowercase{a}F
molecule
}
\begin{figure}
\includegraphics
[width=0.6\linewidth]{
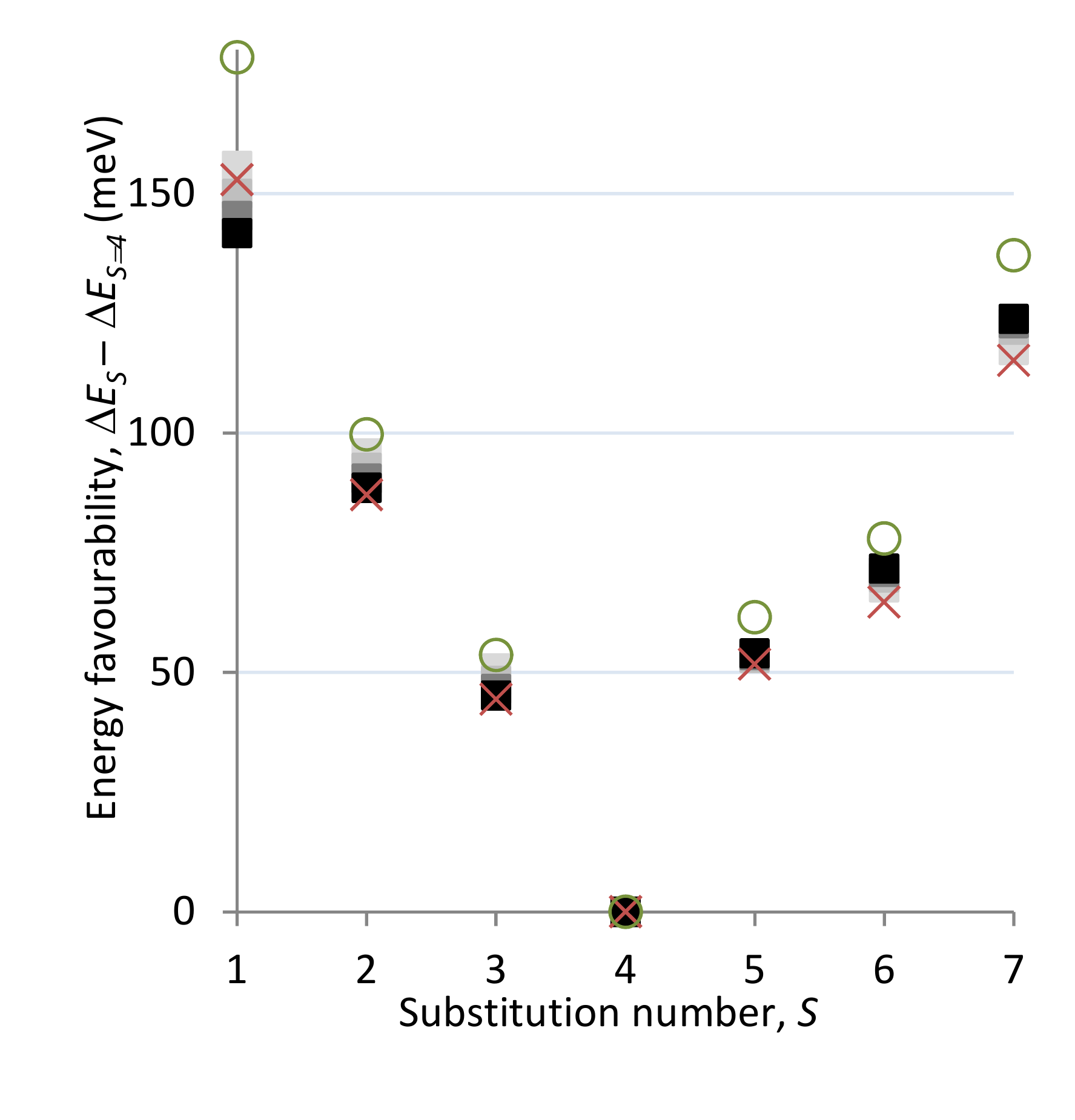}
\caption{
(color online)
Energy favourability
versus
the number
$S$
of
Ar
atoms substituted
for a 
BaF 
molecule.
The data plotted is 
that of 
Table~\ref{table:subs}.
Squares of increasing 
darkness represent the 
simulations 
with increasing numbers
of
Ar 
atoms.
The 
exes
and 
circles 
correspond to
the rows marked
$a$ 
and 
$b$
in 
Table~\ref{table:subs},
and
show the sensitivity 
to the choice of 
BaF-Ar
potential
and 
Ar-Ar
potential,
respectively.
$S=4$
is clearly favoured.
}
\label{fig:subs}
\end{figure}

When embedded in a matrix,
the 
BaF
molecule 
substitutes for 
$S$
argon atoms.
To determine 
which integer
$S$
is the most energetically
favourable,
we compare
values of 
\begin{equation}
\Delta E_{n,m,S}
=
E_{n,m,S}-S E_{\rm coh},
\label{eq:EvsS}
\end{equation}
where 
$E_{n,m,S}$
is calculated  
using 
Eq.~(\ref{eq:totalE})
for 
a sphere of 
radius 
$b_m$
of 
nonfixed
Ar
atoms
(of which 
$S$
are removed and 
replaced with 
a
BaF
molecule)
inside of a  
spherical shell
of fixed 
Ar
atoms
that
extends to 
a radius of 
$b_n$.
The 
$S E_{\rm coh}$
term
corrects for
the missing 
Ar
cohesive
energy 
from the
removal of
$S$
isolated
Ar 
atoms.
The lowest 
value of 
$\Delta E_{n,m,S}$
occurs for 
$S=4$.
This is in contrast 
to the case for a
neutral 
Ba
atom,
where 
$S=6$
is the preferred 
substitution 
\cite{kleshchina2019stable}.
As can be seen in 
Ref.~\cite{koyanagi2022accurate},
the 
F
side of the 
BaF
molecule
bonds more strongly
to the
Ar 
atoms
and this reduces 
the preferred
value of
$S$
for 
BaF 
as 
compared 
to 
Ba.
Table~\ref{table:subs}
and 
Fig.~\ref{fig:subs}
show the values
of 
$
\Delta E_{n,m,S}
-
\Delta E_{n,m,S=4}
$
for various values 
of 
$n$
and
$m$.

\begin{table}
\caption{\label{table:subs} 
Comparison of energetic 
favourability
of a
BaF
molecule substituting for 
$S$
Ar 
atoms
for increasing 
cluster size
$N$
with an
increasing
number 
$M$
of
nonfixed
Ar 
atoms.
To aid in determining
the uncertainties of the 
simulations, 
calculations using 
a less precise 
form of the 
BaF-Ar potential
(extrapolated from 
$n \zeta$,
with
$n=2$,
3
and
4)
are also shown,
as are 
calculations 
for which the 
Ar-Ar
potential is 
not scaled.
$S=4$
is strongly favoured.
}
\begin{tabular}{cccccccccccc}
&&&&
\multicolumn{7}{c}{$
\Delta E_{n,m,S}
-
\Delta E_{n,m,S=4}
$ (meV)}\\
$n$
&
$m$
&
$N_{S=4}$
&
$M_{S=4}$
&
$S$: 1
&
2
&
3
&
4
&
5
&
6
&
7
\\
\hline
13
&
5
&
317
&
75
&
156
&
96
&
51
&
0
&
51
&
68
&
117
\\
19
&
7
&
527
&
131
&
150
&
93
&
48
&
0
&
52
&
70
&
122
\\
19
&
7
&
527
&
131
&
153$^a$
&
87$^a$
&
44$^a$
&
0$^a$
&
52$^a$
&
65$^a$
&
115$^a$
\\
19
&
7
&
527
&
131
&
178$^b$
&
100$^b$
&
54$^b$
&
0$^b$
&
61$^b$
&
78$^b$
&
137$^b$
\\
27
&
11
&
883
&
221 
&
145
&
90
&
47
&
0
&
54
&
71
&
123
\\
47
&
20
&
1957 
&
551
&
142
&
89
&
45
&
0
&
54
&
72
&
124
\\ 
\end{tabular}

$^a$This row uses the less-precise 234 BaF-Ar potential. 
One quarter of the difference between this entry and the 
entry above it gives an estimate of the uncertainty
for the 
previous row
due to uncertainties in the potential calculated
in 
Ref.~\cite{koyanagi2022accurate}.
\\
$^b$This row uses the unscaled Ar-Ar potential. 
The difference between this row and 
the second row
provides a scale for the approximation 
implicit in 
Eq.~(\ref{eq:scaledArAr}).
\end{table}

As can be seen 
from the table,
substituting a 
BaF
molecule 
for 
$S=4$
Ar
atoms 
(i.e., 
a 
tetrasubstitution)
is energetically 
favourable compared to 
other values 
of 
$S$.
There are no local minima
at other values of 
$S$,
which helps to ensure that
the 
BaF
molecules will more
efficiently
move to 
tetrasubstitution 
sites
as the 
BaF-doped
Ar 
solid is 
annealed.
The 
$\gtrsim$50\nobreakdash-meV
energy advantage of the 
tetrasubstitution
is much larger than the 
thermal energy scale 
$k_B T=0.34$~meV
for a doped solid
held at 
4~kelvin. 
As a result, 
it can be expected
that the
BaF 
molecules will
persist in a
tetrasubstitution 
site 
at 
this temperature.

The conclusions drawn from 
Table~\ref{table:subs}
are not affected by 
any of the uncertainties
that we investigated,
as illustrated in 
Fig.\ref{fig:subs}.
From the trend versus
$n$ 
and
$m$
in the table, 
it is clear that the extrapolation
to
even larger clusters will
only lead to corrections
of a few percent.
A recalculation of 
the clusters
using the 
234 
extrapolation 
of
Ref.~\cite{koyanagi2022accurate} 
instead
of the 
345 
extrapolation
indicates 
an uncertainty of 
2\%
or less 
(one quarter of the difference
between row
2
and 
row 
3 
of
Table~\ref{table:subs}).
Repeating 
the simulations 
with 
an 
unscaled 
Ar-Ar
potential
leads to 
corrections 
of approximately
10\%.

\subsection{
Tetrasubstitution
geometry
\label{subSect:geom}}

\begin{figure}
\includegraphics
[width=0.6\linewidth]{
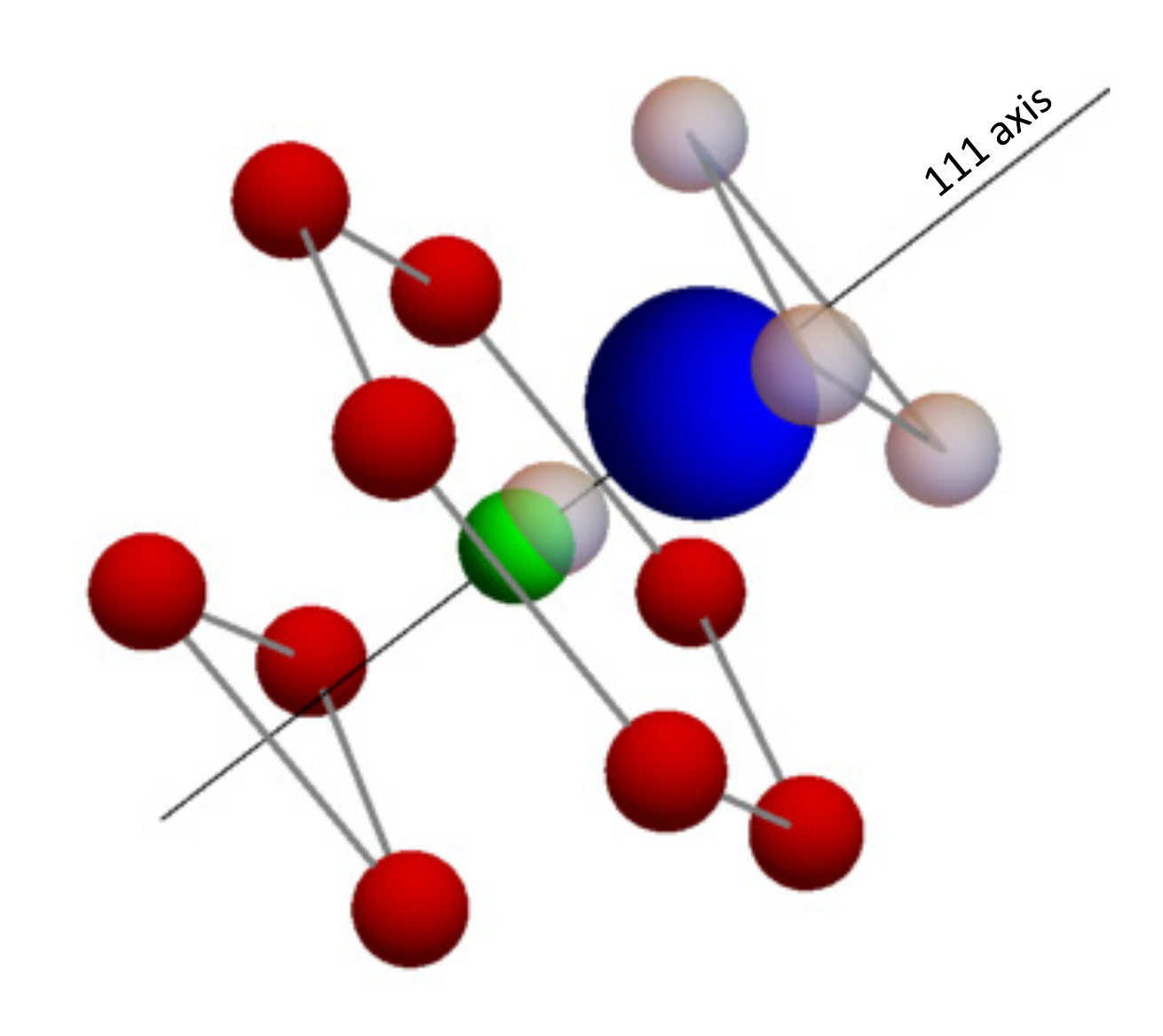}
\caption{
(color online)
The 
BaF
molecule 
substitutes for 
four
Ar 
atoms
(shown in grey)
and is aligned along the 
111
axis 
of the 
Ar
fcc
crystal.
The 
Ba 
atom 
(blue)
is situated 
near the centroid 
of the tetrahedron
defined by the 
four missing 
Ar
atoms,
and the 
F 
atom 
(green)
sits near
the lowest 
missing 
Ar
atom.
The figure shows
the 
remaining 
nine
nearest-neighbour
Ar
atoms
(red)
of the 
F
atom.
The 
Ar atoms are only 
slightly displaced 
from their 
original 
fcc
crystal positions,
as shown in 
Table~\ref{table:displacement}.
}
\label{fig:tetraGeom}
\end{figure}

At its minimum 
energy the favoured
tetrasubstitution 
has the 
BaF
molecule
aligned with the 
111
axis of the 
Ar
fcc
crystal, 
as shown in 
Fig.~\ref{fig:tetraGeom}.
The four missing 
Ar
atoms
(grey in 
the figure)
form a 
tetrahedron,
and the 
Ba
atom 
is situated near 
the centroid of this 
tetrahedron.
The position of the 
BaF
molecule
relative to this 
centroid
is detailed in 
Table~\ref{table:BaFposition}.
Note that the position converges
quickly as the number of 
Ar
atoms in the simulation  
increases.
The basic structure illustrated in 
Fig.~\ref{fig:tetraGeom}
remains the same if the 
less-precise 
234
extrapolation is 
used in the simulation,
as well as if 
the 
unscaled 
form
of the 
Ar-Ar 
potential is used.
In particular,
Table~\ref{table:BaFposition}
shows that 
the position
of the 
BaF
molecule
is not strongly
affected by
the potentials 
used in the calculation.

\begin{table}
\caption{\label{table:BaFposition} 
The position of the 
BaF
molecule relative to the 
centroid 
of the 
tetrahedron 
defined by the 
fcc
positions of the 
four missing 
Ar
atoms 
(grey spheres in  
Fig.~\ref{fig:tetraGeom}).
The positions of the 
Ba 
and 
F 
nuclei
are given,
as is the 
geometric centre
(BaF$_{\rm geom.~cent.}$)
and 
centre of mass
(BaF$_{\rm  c.m.}$) 
of the 
BaF
molecule.
The displacements converge 
quickly with the number of 
Ar atoms used in the simulation.}
\begin{tabular}{ccccccc}
\multicolumn{7}{c}{displacement along axis from centroid (\AA)}
\\
&
$N$: 317
&
527
&
527
&
527
&
883
&
1957 
\\
&
$M$: 75
&
131
&
131
&
131
&
221
&
551
\\
\hline
Ba&0.541&0.538&0.571$^a$&0.512$^b$&0.538&0.537\\
F&2.701&2.698&2.731$^a$&2.572$^b$&2.698&2.697\\
BaF$_{\rm geom.\ cent.}$&1.621&1.618&1.651$^a$&1.592$^b$&1.618&1.617\\
BaF$_{\rm c.m.}$&0.803&0.800&0.833$^a$&0.774$^b$&0.800&0.799\\
\end{tabular}

$^a$This column uses the 
less-precise 
234 
BaF-Ar 
potential. 
One quarter of the 
difference between this entry and the 
entry to its left
gives an estimate of the uncertainty
for the 
previous column
due to uncertainties in 
the potential calculated
in 
Ref.~\cite{koyanagi2022accurate}.
\\
$^b$This column uses the 
unscaled 
Ar-Ar 
potential. 
The difference between this column
and the second column 
gives a scale for the approximation 
implicit in 
Eq.~(\ref{eq:scaledArAr}).
\end{table}

The equilibrium
positions of the 
113
Ar
atoms
nearest to the 
BaF
molecule 
are shown in 
Table~\ref{table:rTheta}.
All of these
Ar 
atoms
are only very slightly 
displaced from their
original 
fcc 
positions,
as shown
in 
Table~\ref{table:displacement}.
The very small displacements
(less than 0.07~\AA;
less than 
2\%
of the 
nearest-neighbour
distance,
$b_1$)
indicate that the 
BaF
molecule
only slightly
perturbs the rest of 
the crystal.
The 
displacements in 
Table~\ref{table:displacement}
are almost independent 
of the potentials 
used and of the number
of
Ar
atoms
used in the 
simulations.

For the 
less-energy-favourable
$S=3$
configuration,
one of the
grey
Ar
atoms from the top
equilateral triangle
of 
Fig.~\ref{fig:tetraGeom}
is present,
and the 
BaF
molecule tilts away from 
the 
111
axis as the 
Ba
atom 
is repelled by 
this additional
Ar atom.
For 
$S=5$,
which is also much
less energy favourable,
the additional
missing
Ar 
atom
is directly 
above the centre
of this equilateral 
triangle, 
and the 
BaF
molecule 
is aligned with the 
111
axis.

\begin{table}
\caption{\label{table:rTheta} 
The positions
($r_{\rm BaF-Ar}$,
$\theta_{\rm BaF-Ar}$)
for the 
113 
nearest
Ar
atoms
relative to the 
BaF
molecule,
where 
($r_{\rm BaF-Ar}$,
$\theta_{\rm BaF-Ar}$)
are relative to the 
midpoint of the 
BaF 
molecule,
as described in 
Section~\ref{sec:methods}.
Also shown
are the number
of 
Ar
atoms
$n_{\rm Ar}$
and 
interaction 
energy
$E_{\rm BaF-Ar}$
per Ar atom
at each 
($r_{\rm BaF-Ar}$, 
$\theta_{\rm BaF-Ar}$).
We include the positions of all of 
these atoms 
in 
Cartesian
coordinates
in the 
Supplementary Materials. 
}
\begin{tabular}{ccccc}
&
$r_{\rm BaF-Ar}$(\AA)
&
$\theta_{\rm BaF-Ar}$($^\circ$)
&
$n_{\rm Ar}$
&
$E_{\rm BaF-Ar}$ (meV)
\\
\hline
1&3.78&79.3&6&-17.11\\
2&4.30&29.9&3&-22.62\\
3&5.02&118.6&3&-0.54\\
4&5.73&49.1&3&-4.46\\
5&5.89&158.2&3&-3.38\\
6&6.17&112.6&6&-6.08\\
7&6.53&84.0&6&-2.96\\
8&6.84&56.8&6&-1.59\\
9&6.93&141.5&3&-5.41\\
10&7.14&17.6&3&-1.32\\
11&7.52&84.8&6&-1.34\\
12&7.93&133.5&6&-2.48\\
13&8.07&32.5&3&-0.58\\
14&8.19&106.9&6&-1.27\\
15&8.52&180.0&1&-2.16\\
16&8.67&64.4&6&-0.41\\
17&8.91&40.1&6&-0.31\\
18&8.97&105.4&3&-0.68\\
19&9.31&156.2&6&-1.08\\
20&9.44&66.6&3&-0.25\\
21&9.52&124.9&6&-0.64\\
22&9.73&104.2&6&-0.39\\
23&9.88&0.0&1&-0.18\\
24&9.95&86.1&12&-0.24\\
\end{tabular}
\end{table}

\begin{table}
\caption{\label{table:displacement} 
The displacement of the 
Ar
atoms 
(see
Table~\ref{table:rTheta}
for numbering)
relative to their 
ideal 
fcc
crystal 
positions.
The parallel
($\parallel$)
and 
perpendicular 
($\perp$)
displacements
are relative to the  
BaF~axis
(the 111 axis),
with positive in the 
direction of 
$\vec{r}_{\rm F}-\vec{r}_{\rm Ba}$
and away from the axis,
respectively.
Note that the displacements
are all 
very small 
(cf. the 
3.756-\AA~Ar 
nearest-neighbour
separation).
The displacements converge 
quickly with the number of 
Ar 
atoms used in the simulation.}
\begin{tabular}{cccccccc}
&
&
\multicolumn{6}{c}{displacement (\AA)}
\\
&
&
$N$: 317
&
527
&
527
&
527
&
883
&
1957 
\\
&
&
$M$: 75
&
131
&
131
&
131
&
221
&
551
\\
\hline
1& $\parallel$ &0.018&0.019&0.020$^a$&0.017$^b$&0.019&0.019\\
& $\perp$ &-0.039&-0.040&-0.047$^a$&-0.031$^b$&-0.042&-0.042\\
2& $\parallel$ &-0.028&-0.027&-0.016$^a$&-0.015$^b$&-0.027&-0.026\\
& $\perp$ &-0.026&-0.027&-0.026$^a$&-0.019$^b$&-0.028&-0.028\\
3& $\parallel$ &-0.009&-0.013&-0.014$^a$&-0.012$^b$&-0.015&-0.015\\
& $\perp$ &0.057&0.064&0.068$^a$&0.064$^b$&0.067&0.068\\
4& $\parallel$ &0.001&0.006&0.007$^a$&0.007$^b$&0.007&0.008\\
& $\perp$ &-0.008&-0.004&-0.005$^a$&-0.003$^b$&-0.005&-0.005\\
5& $\parallel$ &-0.013&-0.015&-0.024$^a$&-0.020$^b$&-0.016&-0.018\\
& $\perp$ &0.016&0.019&0.022$^a$&0.020$^b$&0.020&0.021\\
6& $\parallel$ &0.011&0.012&0.012$^a$&0.010$^b$&0.011&0.012\\
& $\perp$ &-0.037&-0.045&-0.045$^a$&-0.039$^b$&-0.045&-0.046\\
7& $\parallel$ &0.004&0.005&0.005$^a$&0.004$^b$&0.004&0.005\\
& $\perp$ &-0.010&-0.013&-0.014$^a$&-0.009$^b$&-0.012&-0.013\\
8& $\parallel$ &-0.003&-0.003&-0.002$^a$&-0.001$^b$&-0.003&-0.002\\
& $\perp$ &-0.010&-0.013&-0.014$^a$&-0.010$^b$&-0.013&-0.014\\
9& $\parallel$ &0.025&0.024&0.022$^a$&0.021$^b$&0.024&0.023\\
& $\perp$ &-0.021&-0.023&-0.023$^a$&-0.022$^b$&-0.025&-0.026\\
\end{tabular}

$^a$This column uses the 
less-precise 
234 
BaF-Ar 
potential. 
One quarter of the 
difference between this entry and the 
entry to its left
gives an estimate of the uncertainty
for the 
previous column
due to uncertainties in 
the potential calculated
in 
Ref.~\cite{koyanagi2022accurate}.
\\
$^b$This column uses the 
unscaled 
Ar-Ar 
potential. 
The difference between this column
and column two 
gives a scale for the approximation 
implicit in 
Eq.~(\ref{eq:scaledArAr}).
\end{table}

\subsection{
Potential for preventing
BaF rotations and migration}

\begin{figure}
  \includegraphics[width=.6\linewidth]{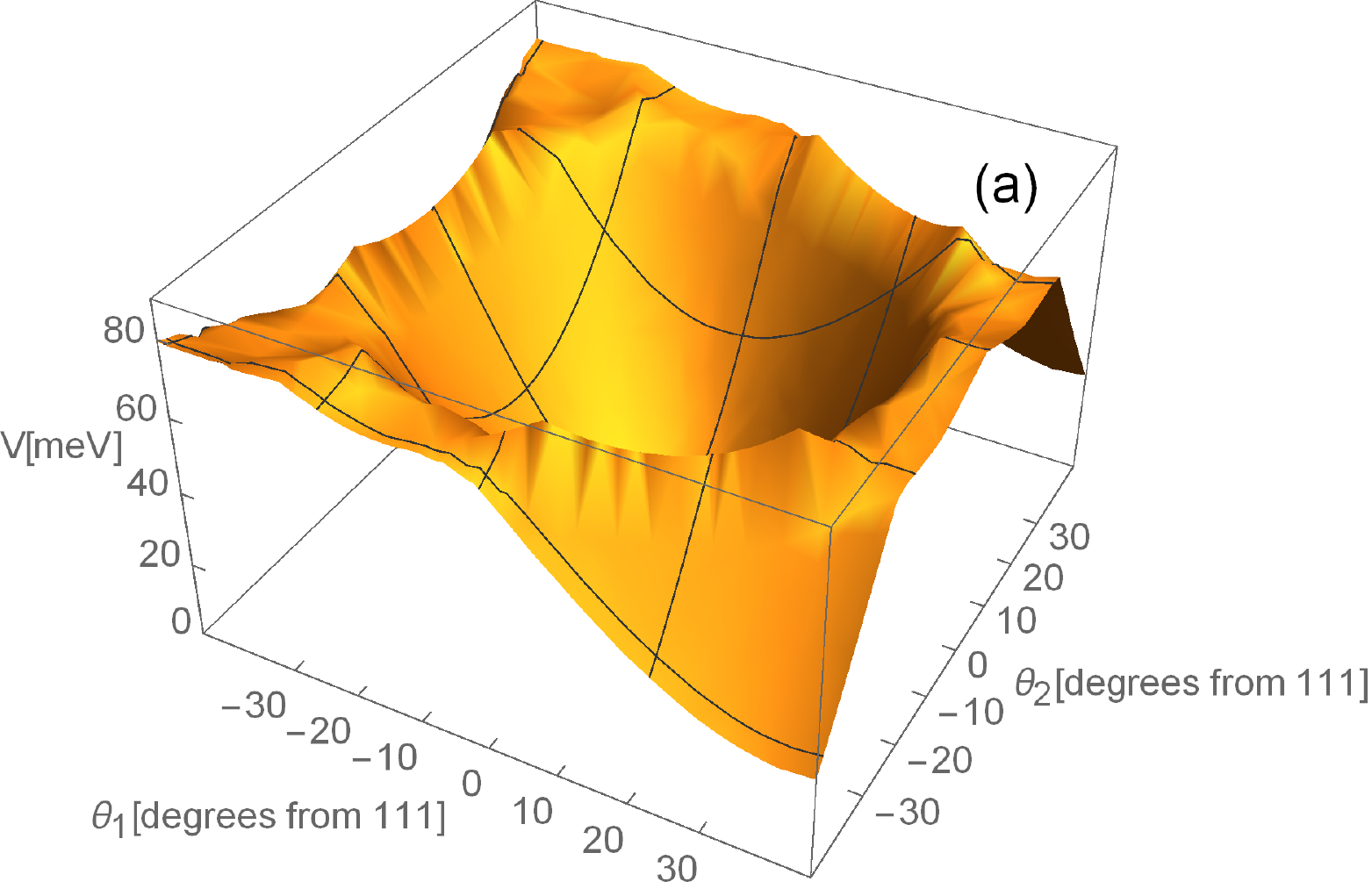}
  \includegraphics[width=.4\linewidth]{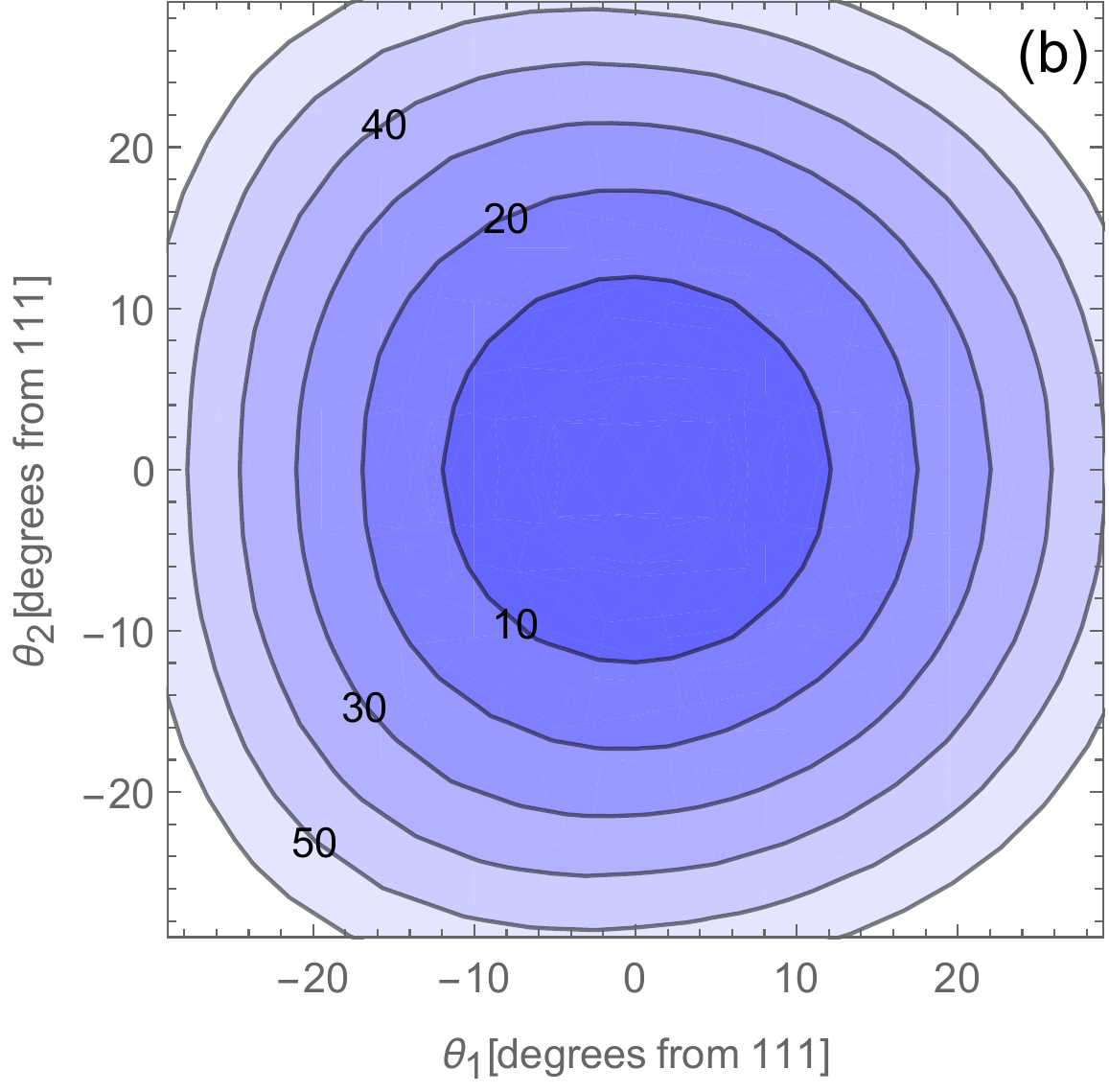}
\caption{
(color online)
A three-dimensional plot
(a) 
and 
a contour plot
(b)
of
$V(\theta_1,\theta_2)$ 
showing the
energy cost 
in meV
for changing the 
orientation of the 
BaF
molecule from
its preferred
111
axis
in two perpendicular 
directions:
$\theta_1$
(towards the (2,-1,-1) direction)
and 
$\theta_2$
(towards the (0,1,-1) direction).
}
\label{fig:angleBarrier}
\end{figure}

The 
energy cost of 
orienting the
BaF
molecule 
away from the 
111
axis of the 
Ar
crystal
is calculated
to determine
the potential
energy barrier
that prevents
the
BaF
molecule
from changing its
orientation.
To determine the energy 
cost for a particular 
orientation,
we find the minimum
energy
for the system
for a large number of
fixed orientations
of the 
BaF
molecule
(that is,
we minimize the 
energy 
with the two 
parameters 
that determine
the orientation
of the molecule
fixed, 
while varying the 
other
$3M+3$
parameters
of 
Section~\ref{sec:methods}
that define the centres of mass 
of the 
Ar 
atoms
and the
BaF
molecule).
The resulting 
potential barrier
is shown in 
Fig.~\ref{fig:angleBarrier}.
From the figure, 
it can be seen that 
a deep
($>$50~meV)
potential
well keeps the 
BaF
molecule aligned along 
the 111 axis.
Identical wells are present
at eight symmetric 
axes: $\pm1\pm1\pm1$.
For a 
4-kelvin
doped solid,
these wells are 
sufficiently deep to 
confine the orientation
to one of these 
eight
orientations.
These fixed orientations,
along with methods to 
separately address 
individual orientations
\cite{vutha2018orientation},
allow for an
eEDM 
measurement
without the need for 
an external electric field,
as is used in all other 
eEDM
measurements 
(see,
e.g., 
Refs.~\cite{acme2018improved},
\cite{Cairncross2017}
and
\cite{hudson2011improved}).
By separately addressing 
oppositely-oriented
molecules  
\cite{vutha2018orientation},
simultaneous 
eEDM
measurements
are planned
using 
these
two
sets of 
interspersed 
molecules.

Similarly,
to determine the 
energy cost for a 
BaF 
molecule being
at a position
away 
from its 
equilibrium 
(see 
Table~\ref{table:BaFposition}),
the simulations are repeated
with the 
BaF
centre-of-mass
position 
fixed
while 
minimizing the energy by 
varying the remaining
$3M+2$
parameters.
This minimization
is repeated for a large number 
of positions.
For positions within
approximately
0.7~\AA~of
the equilibrium,
the 
preferred orientation 
of the 
BaF
molecule
continues to be along the 
111 axis.
A  
displacement
from equilibrium
by 0.7~\AA~
in any direction
already
has an
energy cost of
approximately 
50~meV
or greater,
making 
movement 
by this distance
inaccessible 
for a 
4-kelvin 
doped 
solid.
Having their position
fixed by the matrix will
ensure that the molecules
will not migrate 
through the solid.

As with the other 
results of this 
work, 
the conclusions
that the 
BaF 
molecule
cannot 
reorient
itself
or
migrate
are
not affected by 
any of the uncertainties
discussed 
at the end of 
Section~\ref{sec:methods}.

\section{
Conclusions}
Calculations 
of the local environment 
of a 
BaF
molecule within
an argon matrix
are reported.
It is found that
the molecule
strongly prefers to
replace 
four
Ar
atoms
and 
is aligned along the 
111 
axis of 
the 
argon crystal.
The remaining 
Ar
atoms are 
found to be only 
slightly 
displaced 
from their 
original 
fcc
crystal positions.
The single type of 
strongly-preferred 
site,
the inability of
the molecule
to reorient or 
migrate at 
a temperature of
4 kelvin 
and 
the small perturbation
of the rest of the 
Ar
crystal 
are important 
features
for the 
planned 
eEDM
measurement 
by the 
EDM$^3$
collaboration 
using 
matrix-isolated
BaF
molecules.

\section*{Acknowledgements}
This work is supported by
the
Alfred P. Sloan Foundation,
the 
Gordon and Betty Moore Foundation,
the 
Templeton Foundation
in conjunction with the 
Northwestern Center for Fundamental Physics,
the
Natural Sciences and Engineering Research Council
of Canada
and
York University.
Computations for this work 
required more than 
15
core-years 
of 
CPU
time
and 
were 
enabled by support provided by 
Compute Canada.

\section*{Disclosure Statement}
The authors report there are no competing interests to declare.

\newpage
\setcounter{table}{0}
\renewcommand{\thetable}{S\arabic{table}}
\begin{table}
\caption{
\textbf{Supplementary Materials:}
Cartesian coordinates of atoms for the 
$S=4$.
}
\begin{tabular}{rrrrrrrrrrrr}
atom&x(\AA)&y(\AA)&z(\AA)&atom&x(\AA)&y(\AA)&z(\AA)&atom&x(\AA)&y(\AA)&z(\AA)\\
Ba&-1.02&-1.02&-1.02&
F&0.23&0.23&0.23\\
Ar 1&-2.62&0.01&2.64&
Ar 2&-2.62&2.64&0.01&
Ar 3&0.01&-2.62&2.64\\
Ar 4&0.01&2.64&-2.62&
Ar 5&0.01&2.63&2.63&
Ar 6&2.64&-2.62&0.01\\
Ar 7&2.64&0.01&-2.61&
Ar 8&2.63&0.01&2.63&
Ar 9&2.63&2.63&0.01\\
Ar 10&-5.38&0.02&0.02&
Ar 11&0.02&-5.38&0.02&
Ar 12&0.02&0.02&-5.38\\
Ar 13&0.01&0.01&5.31&
Ar 14&0.01&5.31&0.01&
Ar 15&5.31&0.01&0.01\\
Ar 16&-5.34&-2.66&-2.66&
Ar 17&-5.28&-2.64&2.63&
Ar 18&-5.28&2.63&-2.64\\
Ar 19&-5.30&2.65&2.65&
Ar 20&-2.66&-5.34&-2.66&
Ar 21&-2.64&-5.28&2.63\\
Ar 22&-2.66&-2.66&-5.34&
Ar 23&-2.65&-2.65&5.31&
Ar 24&-2.64&2.63&-5.28\\
Ar 25&-2.65&2.65&5.30&
Ar 26&-2.65&5.31&-2.65&
Ar 27&-2.65&5.30&2.65\\
Ar 28&2.63&-5.28&-2.64&
Ar 29&2.65&-5.30&2.65&
Ar 30&2.63&-2.64&-5.28\\
Ar 31&2.65&-2.65&5.30&
Ar 32&2.65&2.65&-5.30&
Ar 33&2.65&2.65&5.30\\
Ar 34&2.65&5.30&-2.65&
Ar 35&2.65&5.30&2.65&
Ar 36&5.31&-2.65&-2.65\\
Ar 37&5.30&-2.65&2.65&
Ar 38&5.30&2.65&-2.65&
Ar 39&5.30&2.65&2.65\\
Ar 40&-5.29&-5.29&-0.01&
Ar 41&-5.29&-0.01&-5.29&
Ar 42&-5.30&0.00&5.30\\
Ar 43&-5.30&5.30&0.00&
Ar 44&-0.01&-5.29&-5.29&
Ar 45&0.00&-5.30&5.30\\
Ar 46&0.00&5.30&-5.30&
Ar 47&0.00&5.30&5.30&
Ar 48&5.30&-5.30&0.00\\
Ar 49&5.30&0.00&-5.30&
Ar 50&5.30&0.00&5.30&
Ar 51&5.30&5.30&0.00\\
Ar 52&-7.98&-2.66&0.00&
Ar 53&-7.98&0.00&-2.66&
Ar 54&-7.98&0.00&2.66\\
Ar 55&-7.98&2.66&0.00&
Ar 56&-2.66&-7.98&0.00&
Ar 57&-2.66&0.00&-7.98\\
Ar 58&-2.65&0.00&7.96&
Ar 59&-2.65&7.96&0.00&
Ar 60&0.00&-7.98&-2.66\\
Ar 61&0.00&-7.98&2.67&
Ar 62&0.00&-2.66&-7.98&
Ar 63&0.00&-2.65&7.96\\
Ar 64&0.00&2.66&-7.98&
Ar 65&0.00&2.66&7.97&
Ar 66&0.00&7.96&-2.65\\
Ar 67&0.00&7.97&2.66&
Ar 68&2.66&-7.98&0.00&
Ar 69&2.67&0.00&-7.98\\
Ar 70&2.66&0.00&7.97&
Ar 71&2.66&7.97&0.00&
Ar 72&7.96&-2.65&0.00\\
Ar 73&7.96&0.00&-2.65&
Ar 74&7.97&0.00&2.66&
Ar 75&7.97&2.66&0.00\\
Ar 76&-5.31&-5.31&-5.31&
Ar 77&-5.30&-5.30&5.30&
Ar 78&-5.30&5.30&-5.30\\
Ar 79&-5.31&5.31&5.31&
Ar 80&5.30&-5.30&-5.30&
Ar 81&5.31&-5.31&5.31\\
Ar 82&5.31&5.31&-5.31&
Ar 83&5.31&5.31&5.31&
Ar 84&-7.97&-5.31&-2.65\\
Ar 85&-7.96&-5.30&2.65&
Ar 86&-7.97&-2.65&-5.31&
Ar 87&-7.96&-2.65&5.30\\
Ar 88&-7.96&2.65&-5.30&
Ar 89&-7.96&2.66&5.31&
Ar 90&-7.96&5.30&-2.65\\
Ar 91&-7.96&5.31&2.66&
Ar 92&-5.31&-7.97&-2.65&
Ar 93&-5.30&-7.96&2.65\\
Ar 94&-5.31&-2.65&-7.97&
Ar 95&-5.31&-2.65&7.96&
Ar 96&-5.30&2.65&-7.96\\
Ar 97&-5.31&2.65&7.96&
Ar 98&-5.31&7.96&-2.65&
Ar 99&-5.31&7.96&2.65\\
Ar 100&-2.65&-7.97&-5.31&
Ar 101&-2.65&-7.96&5.30&
Ar 102&-2.65&-5.31&-7.97\\
Ar 103&-2.65&-5.31&7.96&
Ar 104&-2.65&5.30&-7.96&
Ar 105&-2.65&5.31&7.96\\
Ar 106&-2.65&7.96&-5.31&
Ar 107&-2.65&7.96&5.31&
Ar 108&2.65&-7.96&-5.30\\
Ar 109&2.66&-7.96&5.31&
Ar 110&2.65&-5.30&-7.96&
Ar 111&2.65&-5.31&7.96\\
Ar 112&2.66&5.31&-7.96&
Ar 113&2.65&5.31&7.96&
Ar 114&2.65&7.96&-5.31\\
Ar 115&2.65&7.96&5.31&
Ar 116&5.30&-7.96&-2.65&
Ar 117&5.31&-7.96&2.66\\
Ar 118&5.30&-2.65&-7.96&
Ar 119&5.31&-2.65&7.96&
Ar 120&5.31&2.66&-7.96\\
Ar 121&5.31&2.65&7.96&
Ar 122&5.31&7.96&-2.65&
Ar 123&5.31&7.96&2.65\\
Ar 124&7.96&-5.31&-2.65&
Ar 125&7.96&-5.31&2.65&
Ar 126&7.96&-2.65&-5.31\\
Ar 127&7.96&-2.65&5.31&
Ar 128&7.96&2.65&-5.31&
Ar 129&7.96&2.65&5.31\\
Ar 130&7.96&5.31&-2.65&
Ar 131&7.96&5.31&2.65&
\end{tabular}
\end{table}

\end{document}